\documentclass[prd,nofootinbib,preprint]{revtex4}

\usepackage{amsmath,amssymb}
\usepackage{epsfig}

\newcommand{\sla}[1]{/\!\!\!#1}

\begin{document}

\preprint{\hbox{YITP-SB-22-04}} 

\title{Probing Trilinear Gauge Boson Interactions via Single
  Electroweak Gauge Boson Production at the LHC}

\author{O.\ J.\ P.\ \'Eboli \footnote{E-mail: eboli@fma.if.usp.br}}
\affiliation{Instituto de F\'{\i}sica da USP, C.P. 66.318, S\~ao
  Paulo, SP 05315-970, Brazil.}

\author{M.C. Gonzalez-Garcia \footnote{E-mail:
    concha@insti.physics.sunysb.edu}}
\affiliation{ Y.I.T.P., SUNY at Stony Brook, Stony Brook, NY 11794-3840, USA\\
  IFIC, Universitat de Val\`encia - C.S.I.C., Apt 22085, 46071
  Val\`encia, Spain}


\begin{abstract}
\vspace*{1cm}

We analyze the potential of the CERN Large Hadron Collider (LHC) to study
anomalous trilinear vector--boson interactions $W^+ W^- \gamma$ and $W^+ W^-
Z$ through the single production of electroweak gauge bosons via the weak
boson fusion processes $ q q\to q q W (\to \ell^\pm \nu)$ and $ q q \to q q
Z(\to \ell^+ \ell^-)$ with $\ell = e$ or $\mu$.  After a careful study of the
standard model backgrounds, we show that the single production of electroweak
bosons at the LHC can provide stringent tests on deviations of these vertices
from the standard model prediction. In particular, we show that single gauge
boson production exhibits a sensitivity to the couplings $\Delta
\kappa_{Z,\gamma}$ similar to that attainable from the analysis of electroweak
boson pair production.

\end{abstract}

\maketitle

\section{Introduction}

Within the framework of the SM, the structure of the trilinear and quartic
vector--boson couplings is completely determined by the $SU(2)_L \times
U(1)_Y$ gauge symmetry. The study of these interactions can either lead to an
additional confirmation of the model or give some hint on the existence of new
phenomena at a higher scale~\cite{anomalous}. The triple gauge--boson vertices
(TGV's) have been probed directly at the Tevatron~\cite{teva} and
LEP~\cite{lep,ewwg} through the production of vector--boson pairs and the
experimental results agree with the SM predictions within ${\cal O}$(10\%);
see Table \ref{tab:outros}. Moreover, TGV's contribute at the one--loop level
to the $Z$ physics and consequently they can also be indirectly constrained by
precision electroweak data~\cite{LEPI}.  At the LHC, the TGV's will be subject
to a more severe scrutiny via the production of electroweak boson pairs ($W
\gamma$ and $WZ$) \cite{ubdz} which will probe these couplings at the few
percentage level \cite{tri@lhc}.

In this work we analyze the LHC potential to study the TGV's through the weak
boson fusion (WBF) reactions
\begin{eqnarray}
p\,p \,\rightarrow\, j\; j\; W^+\,\rightarrow \, j\; j\; l^+\; \nu_l 
\;\; ,
\nonumber  \\
p\,p \,\rightarrow\, j\; j\; W^- \,\rightarrow\,  j\; j\; l^-\;\overline 
{\nu_l} \;\; ,
\label{vjj} \\
p\,p \,\rightarrow\, j\; j\; Z \,\rightarrow\,  j\; j\; l^+ \; l^- \;\; ,
\nonumber 
\end{eqnarray}
with $l=e,\mu$.  These processes are complementary to the electroweak gauge
boson pair production in the analysis of the $WW\gamma$ and $WWZ$ vertices;
certainly a larger number of basic processes and observables can contribute to
a better scrutiny of the TGV's. While $WZ$, $W^+ W^-$, and $W\gamma$
production at the LHC probe the TGV's for time-like momenta of all vector
bosons, the single $W$ and $Z$ productions via WBF presents two electroweak
gauge bosons with space-like momentum transfer.

TGV's in single gauge boson production processes were studied in
Ref.~\cite{ssc} for the SSC energy but had not been discussed in the context
of the LHC. A potential drawback of the single gauge boson production as a
test of electroweak vertices at the LHC energies is the large expected
background from higher order QCD corrections to the Drell--Yan
process\footnote{ Conversely QCD corrections to the electroweak production
  have been shown to be modest ~\cite{oleari}.}.  However, it has been
recently proved in the case of Higgs production~\cite{dieter} that these
backgrounds are under control due to the presence of two very energetic
forward jets.  Furthermore, its theoretical uncertainty can be efficiently
reduced by making use of a calibration region where the backgrounds can be
estimated from data. In this work we show that, indeed, these conclusions
apply to the study of TGV's in single gauge boson production.

In the following we describe the $W^+ W^- V$ ($V=Z$ or $\gamma$) vertices in
terms of the standard Lorentz invariant and CP conserving parameterization,
which is given by the effective Lagrangian \cite{old:wwv}
\begin{eqnarray}
{\cal L}^{\text{WWV}}_{\text{eff}} =&&  - i g_{\text{WWV}}
\left [
g^V_1 ( W^+_{\mu\nu} W^{-\mu} - W^-_{\mu\nu} W^{+\mu} ) V^\nu \right.
\nonumber \\
&& + \kappa_V W^+_\mu W^-_\nu V^{\mu\nu} +
\frac{\lambda_V}{M_W^2} W_\mu^{+\nu} W_\nu^{-\rho} V_\rho^\mu   
\nonumber \\
&& \left. - i g^V_5 \epsilon^{\mu\nu\rho\sigma}
(W^+_\mu \partial_\rho W^-_\nu
- W^-_\nu \partial_\rho W^+_\mu ) V_\sigma  \right ] \;\; ,
\label{WWV}
\end{eqnarray}
where $V_{\mu\nu} = \partial_\mu V_\nu - \partial_\nu V_\mu$, $g_{WW\gamma} =
e$, and $g_{\text{WWZ}} = e c_W/s_W$, with $s_W (c_W) =\sin (\cos) \theta_W$.
The first three terms in Eq.~(\ref{WWV}) are C and P invariant while the last
one violates both C and P.  Electromagnetic gauge invariance implies that $1 -
g_1^\gamma = g_5^\gamma =0$.  Within the framework of the SM, $g_1^\gamma =
g_1^Z = \kappa_\gamma = \kappa_Z = 1$ and $\lambda_\gamma = \lambda_Z = g^Z_5
=0$.

Since the standard model is consistent with the available experimental data,
it is natural to parameterize the anomalous TGV's in terms of an effective
Lagrangian which exhibits the $SU(2)_L\times U(1)_Y$ gauge invariance. The
particular way this symmetry is realized depends on the particle content at
low energies. If the spectrum at low energies does not exhibit a light Higgs
boson, this symmetry has to be non-linearly realized and the triple gauge
boson vertex can be parameterized as Eq.~(\ref{WWV}) with the couplings
$g_1^Z$ , $\kappa_\gamma$, $\kappa_Z$, $\lambda_\gamma$, $\lambda_Z$, and
$g^Z_5$ being independent parameters~\cite{london}.

Conversely, if a light Higgs boson is present, the symmetry can be realized
linearly \cite{buch,deruj,hisz}. In this case the leading effects of new
interactions are described by eleven dimension--6 operators ${\cal O}_i$
\begin{equation}
{\cal L}^{\text{linear}}_{\text{eff}}  = 
\sum_i \frac{f_i}{\Lambda^{2}} {\cal O}_i  \;\; ,
\end{equation}
at energies below the new physics scale $\Lambda$. Three of these
operators \cite{deruj}, namely,
\begin{eqnarray}
{\cal O}_{B}   &=& (D_\mu \Phi)^\dagger
\hat B^{\mu\nu} (D_\nu \Phi) \;\; , 
\nonumber \\
{\cal O}_{W} &=& (D_\mu \Phi)^\dagger
\hat W^{\mu\nu} (D_\nu \Phi)  \;\; ,
\label{lin:op}
\\ 
{\cal O}_{WWW} &=& {\text{Tr}} \left [\hat W_{\mu\nu}
\hat W^{\nu\rho}\hat W_\rho^{\mu} \right ] \;\; , 
\nonumber
\end{eqnarray}
modify the triple gauge boson couplings without affecting the gauge boson
two--point functions at tree level; the so called ``blind'' operators. In our
notation, $\hat{B}_{\mu\nu} = i (g^\prime/2) B_{\mu\nu}$ and $\hat{W}_{\mu\nu}
= i (g/2) \sigma^a W^a_{\mu\nu}$ with $B_{\mu\nu}$ and $W^a_{\mu\nu}$ being
the $U(1)_Y$ and $SU(2)_L$ full field strengths and $\sigma^a$ representing
the Pauli matrices.  In this framework, it is expected that $g^Z_5$ should be
suppressed since it is related to a dimension 8 operator \cite{our:epsb}.

The anomalous couplings of the parameterization (\ref{WWV}) are related to the
coefficients of the linearly realized effective Lagrangian by
\begin{eqnarray}
\Delta g^Z_1 &=& f_W~\frac{m_Z^2}{2\Lambda^2} \;\; ,
\label{lin:1} \\
\Delta \kappa_Z &=& [f_W-s_W^2 (f_B+f_W)]~\frac{m_Z^2}{2\Lambda^2} \;\; ,
\label{lin:2} \\
\lambda_Z &=& f_{WWW}~ \frac{3m_W^2 g^2}{2\Lambda^2} \;\; . 
\label{lin:3}
\end{eqnarray}
It is interesting to notice that these effective operators lead to the
following relation between the coefficients of Lagrangian (\ref{WWV}),
defining the HISZ scenario \cite{hisz}:
\begin{eqnarray}
        \Delta \kappa_\gamma &=& \frac{c_W^2}{s_W^2} 
        \left( \Delta g_1^Z - \Delta \kappa_Z  \right)
        \;\; , \label{rel1}
\\
        \lambda_\gamma &=& \lambda_Z \;\; .
\label{rel}
\end{eqnarray}

\section{Calculational tools}

We are considering the production of electroweak gauge bosons $W^\pm$ and $Z$
in WBF, $qq\to qqV^*V^* \to qqV$ ($V=W$ or $Z$), with subsequent decays $Z \to
\ell^+ \ell^-$ or $W^\pm \to \ell^\pm \nu_\ell$ with $\ell = e$ or $\mu$. The
signal is thus characterized by two quark jets, which typically enter in the
forward and backward regions of the detector and are widely separated in
pseudo-rapidity, and by two charged leptons or a charged lepton accompanied by
a large transverse momentum imbalance. Significant backgrounds to the
anomalous signal arise from $Zjj$ and $Wjj$ production, which can take place
via standard electroweak subprocesses (including both the WBF and the emission
of the electroweak gauge boson from a quark line) and most copiously through
Drell--Yan gauge boson production production associated with further real
emission. Another potential background is the QCD production of top quark
pairs, with at least one top quark decaying semileptonicaly.

The signal and backgrounds were simulated at the parton level with full tree
level matrix elements.  This was accomplished by numerically evaluating
helicity amplitudes for all subprocesses. Backgrounds include all order
$\alpha_s^2$ real emission corrections to Drell--Yan production, to be called
QCD $Wjj$ and $Zjj$ processes, and cross sections are calculated with code
based on Ref.~\cite{BHOZ}.  The second large class of processes are the
anomalous signal and the electroweak background; this last one denoted by EW
$Wjj$ and $Zjj$ production.  The code for these processes is based on
Ref.~\cite{CZ}.  Madgraph \cite{mad} code was also used to simulate the QCD $t
\bar{t}$ background at tree level. For all QCD effects, the running of the
strong coupling constant is evaluated at one--loop order, with $\alpha_s(M_Z)
= 0.12$. We employed CTEQ5L parton distribution functions \cite{CTEQ5_pdf}
throughout. We took the electroweak parameters $\sin^2 \theta_W = 0.23124$,
$\alpha_{em} = 1/128.93$, $m_Z = 91.189$ GeV, and $m_W = 79.95$ GeV, which was
obtained imposing the tree level relation $\cos \theta_W = m_W/m_Z$.  We
simulate experimental resolutions by smearing the energies (but not
directions) of all final state partons with a Gaussian error given by
$\Delta(E)/E = 0.5/\sqrt{E} \oplus 0.02$ ($E$ in GeV) while for charged
leptons we used a resolution $\Delta(E)/E = 0.02/\sqrt{E}$.

An important feature of the WBF signal is the absence of color exchange
between the final state quarks, which leads to a depletion of gluon emission
in the region between the two tagging jets. We can enhance the signal to
background ratio by vetoing additional soft jet activity in the central region
\cite{veto}. A central jet veto is ineffective against the EW $Wjj$ and $Zjj$
backgrounds which possess the same color structure as the signal.  For the QCD
backgrounds, however, there is color exchange in the $t$--channel and
consequently a more abundant production of soft jets, with $p_T>20$~GeV, in
the central region \cite{CZ}. The probability of an event to survive such a
central jet veto has been analyzed for various processes in
Ref.~\cite{rainth}, from which we take the veto survival probabilities
\begin{equation}
P_{\rm surv}^{\rm EW}=0.82 \;\;\;\;\;\; , \;\;\;\;\;\;
 P_{\rm surv}^{\rm QCD}=0.28 \;\; ,
\label{eq:psurv}
\end{equation}
which are appropriate for the hard tagging jet cuts to be used below.

It is important to note that the operators in Eq.~(\ref{WWV}) lead to
tree--level unitarity violation in $2\to2$ processes at high energies.  The
standard procedure to avoid this unphysical behavior of the cross section and
to obtain meaningful limits is to multiply the anomalous couplings
($g^i_{\text{ano}}$) by a form factor
\begin{equation}
g_{\text{ano}}^i 
\rightarrow \frac{g_{\text{ano}}^i}{
\left [
  \left (1+\frac{|q^2_1|}{\Lambda^2} \right) 
  \left (1+\frac{|q^2_2|}{\Lambda^2}\right ) 
  \left (1+\frac{|q^2_3|}{\Lambda^2}\right ) 
\right]^n} 
\label{formfactor}
\end{equation}
with $q_i$ standing for the four-momenta of the gauge bosons in the
vertex. In our analysis we chose $\Lambda=2.5$ TeV and $n=1$.
At $e^+e^-$ colliders the center--of--mass energy is fixed and the
introduction of the form factor (\ref{formfactor}) is basically equivalent to
a rescaling of the anomalous couplings, therefore we should perform this
rescaling when comparing results obtained at hadron and $e^+e^-$ colliders.
For example, the LEP are weakened by a factor $\simeq 1$\% for our choice of
$n$ and $\Lambda$.

Altogether the cross sections for processes (\ref{vjj}) can be written
as
 \begin{equation}
\sigma=\sigma_{\text{sm}}+ \sum_{i=1}^6\sigma^{\text{int}}_i g^i_{\text{ano}} 
+\sum_{i=1}^6\sum_{j \geq i}\sigma^{\text{ano}}_{ij} g_{\text{ano}}^i
g_{\text{ano}}^j  \;\; , 
\label{def:sigma}
\end{equation}
where $\sigma_{\text{sm}}$, $\sigma^{\text{int}}_i$, and
$\sigma^{\text{ano}}_{ij}$ are, respectively, the SM cross section,
interference between the SM and the anomalous contribution, and the pure
anomalous contributions, which contain the interference between the different
TGV's contributions.

\section{Signal and  Background Properties}

The main features of the production of a single electroweak gauge boson via
WBF are the presence of two very energetic forward jets and one or two
isolated charged leptons.  Therefore, we initially imposed the following jet
tagging cuts
\begin{eqnarray}
&& p_T^j > 40 \hbox{ GeV} \;\;\;\;\;\;\; , \;\;\;\; | y_j | < 5.0 \;\; ,
\nonumber
\\
&& | y_{j1} - y_{j2} | > 4.4 \;\;\; , \;\;\;   
y_{j1} \cdot y_{j2} < 0 \;\; ,
\label{cuts1}
\end{eqnarray}
and lepton acceptance and isolation cuts
\begin{eqnarray}
|y_\ell|\leq 2.5 \;\;\; &,& \;\;\;
p_{T}^\ell \geq 20\; {\rm GeV} \;\; , \label{basiccuts}
\\
\Delta R_{\ell j}\geq 0.6 \;\;\; &,& \;\;\;
\Delta R_{\ell \ell}\geq 0.6 \;\; .\nonumber 
\end{eqnarray}

\begin{figure}[!t]
\centerline{\psfig{figure=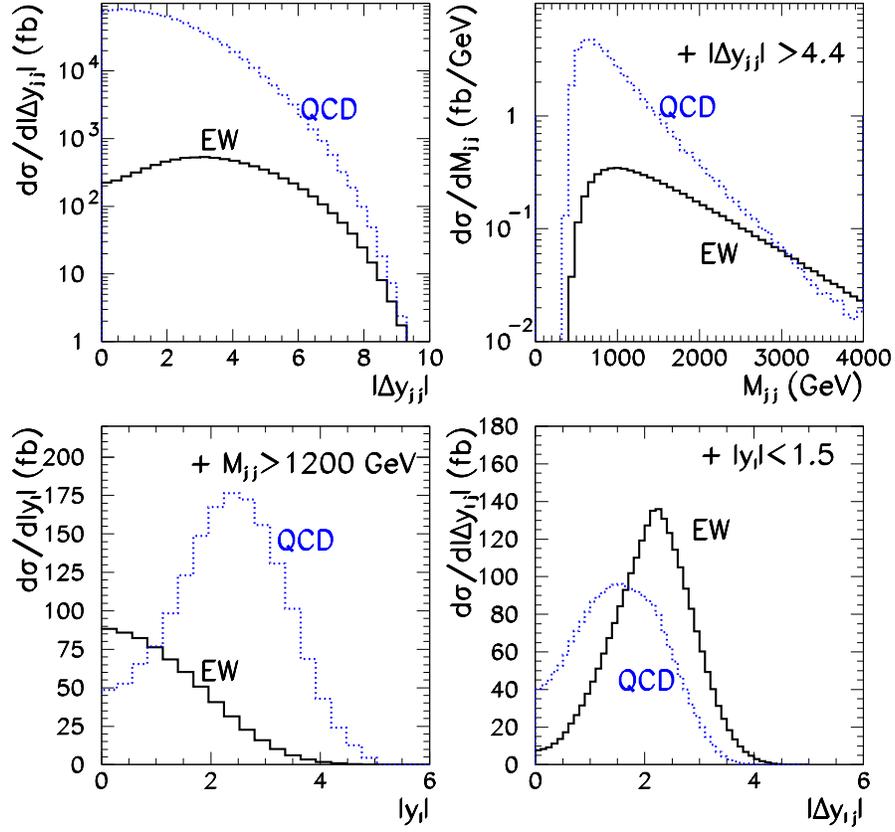,width=5in}} 
\caption{Event distributions for EW and QCD contributions to
  $W^+ jj$ with $W^+\rightarrow l^+\nu_l$ at the LHC. In the left upper panel
  only the cuts in Eqs.~(\ref{cuts1}) and (\ref{basiccuts}) have been included
  with the exception of the separation cut $ | y_{j1} - y_{j2} | > 4.4$. The
  right upper, left lower, and right lower panels illustrate the effect of the
  jet-jet invariant mass cut, the lepton rapidity cut, and the rapidity
  separation between lepton--jet cut respectively in reducing the QCD
  background. In these figures the gap survival probabilities in
  Eq.~(\ref{eq:psurv}) are included except for the left upper panel.}
\label{fig:distbck}
\end{figure}

The effect of the jet rapidity separation cut in suppressing the QCD
background is illustrated in the upper panels in Fig.~\ref{fig:distbck} We
display in the upper left panel of Fig.\ \ref{fig:distbck}, the rapidity
separation between the tagging jets in $W^+ jj$ production ( $ | y_{j1} -
y_{j2} |$) after the cuts (\ref{cuts1}) and (\ref{basiccuts}) prior to the $|
y_{j1} - y_{j2} | > 4.4$ cut for both EW and QCD backgrounds; for most
variables the anomalous signal presents kinematics distributions similar to
the EW background, and consequently, we only show the EW distribution in these
cases. As we can see, the rapidity separation between the tagging jets for the
QCD processes peaks at small values while the EW and signal processes leads to
larger rapidity separations. Consequently imposing the rapidity separation cut
enhances the EW/QCD ratio by a factor $\sim$ 20. However, further cuts are
necessary to reduce the QCD background to acceptable levels.  This can be
achieved by making use of the differences in the invariant mass distribution of
the tagging jets and in the lepton rapidity distributions between the QCD and
EW backgrounds, as illustrated in the upper-right and lower panels of Fig.\ 
\ref{fig:distbck}. First, since the QCD distribution exhibits a larger slope
than the EW and signal ones in the invariant mass distribution of the tagging
jets ($M_{jj}$), a hard cut in $M_{jj}$ tends to suppress the QCD background
and enhance the signal. Second, as expected, the charged lepton rapidity is
larger for the QCD processes since in this case the $W^+$ production is
dominated by the bremsstrahlung of the gauge boson off initial and final state
quarks, and consequently, the charged lepton has a tendency to be closer to
the beam pipe or to the tagging jets.

Taking into account all these  properties of the signal and
backgrounds, we further required
\begin{equation}
|\Delta y_{\ell j}|\ge 2 \;\;\;\;\; , \;\;\;\;\;
|y_\ell |\leq 1.5 \;\;\;\;\; , \;\;\;\;\;
M_{jj}\geq 
\left \{
\begin{array}{ll}
2000\; {\rm GeV} & \hbox{ for $W^\pm$ production}
\\
1200\; {\rm GeV} & \hbox{ for $Z$ production}
\end{array}
\right .
\label{eq:cutsqcd}
\end{equation}
in order to suppress the QCD background.

With these cuts we have selected the phase space region where WBF processes
dominate but so far we have not made any selective cut to discriminate between
the anomalous signal and the SM contribution to WBF.  In order to do so we
make use of the fact that due to the presence of operators with higher
derivatives, and the lost of unitarity, anomalous couplings lead to the
enhancement of the transverse momentum distribution of the electroweak bosons
at high $p_T$'s as illustrated in Fig.~\ref{fig:distptw}. Therefore, the
anomalous signal can be enhanced requiring that
\begin{equation}
p_T^W\geq 300\; {\rm GeV} \;\;\;\;\;\; , \;\;\;\;\;\;
p_T^Z\geq 100\; {\rm GeV} \;\; .
\label{cutano}
\end{equation}
One must note, however, that this enhancement must be effectively cut--off
before it leads to unacceptable violations of unitarity.  As discussed in the
previous section, we follow the standard procedure to avoid this unphysical
behavior of the cross section and multiply the anomalous couplings by a form
factor which we chose to be as Eq.~(\ref{formfactor}).  The effect of this
form factor can be seen comparing the right and left panels of
Fig.~\ref{fig:distptw}. It is worth commenting that pair production of gauge
bosons is affected by the details of the form factors in a different way and
it is probably more sensitive to them~\cite{ssc}. This further stresses the
importance of studying both single and double pair production to obtain the
most meaningful information on the TGV's.

The final results on the EW and QCD background cross section as well as the
anomalous contributions to the coefficients in Eq.~(\ref{def:sigma}) after
applying the cuts (\ref{cuts1})--(\ref{cutano}) is presented in
Tables~\ref{tab:sigmaw+}, ~\ref{tab:sigmaw-}, and \ref{tab:sigmaz}, which
already include the effect of veto survival probability and the detection
efficiency 0.85 for each charged lepton.  From the tables we read that after
applying the cuts (\ref{cuts1})--(\ref{cutano}) about 15/85\% (30/70\%) of the
background $W^\pm$ ($Z$) events are due to QCD/EW processes.  Moreover, we
have verified that the $t\bar t$+n jets background is negligible after
applying cuts ~(\ref{cuts1}) and (\ref{basiccuts}) and vetoing extra jets or
leptons in the central rapidity region of the detector.

\section{Predicting the background}

In this work we estimate the LHC potential to constrain anomalous TGV's in
$jj\ell^\pm \sla{p_T}$ and $jj\ell^+ \ell^-$ events by considering only the
total cross section after cuts, that is, we analyze the sensitivity for TGV's
of a counting experiment.  The sensitivity of this search is thus determined by
the precision with which the background rate in the search region can be
predicted.

This is a challenging task, in particular for the QCD background.  Since the
signal selection is demanding, including double forward jet tagging and
central jet vetoing techniques whose acceptance cannot be calculated with
sufficient precision in perturbative QCD, the theoretically predicted
background can vary up to a factor 3 depending on the choices of factorization
and renormalization scales. Therefore, the possibility of obtaining meaningful
information on deviations from the TGV's is directly limited by our ability to
determine the background directly from LHC data.

This background normalization error can be reduced by relaxing some of the
cuts, {\em i.e.} by considering a larger phase space region as a calibration
region. The background expected in the signal region is then obtained by
extrapolation of the measured events in the calibration region to the signal
region according to perturbative QCD.  This procedure introduces also an
uncertainty, which we denote as QCD--extrapolation uncertainty, due to the
extrapolation to the signal region. However, as we will show, these
uncertainties are smaller than the uncertainties for the total cross section.

We defined the calibration region by the cuts
(\ref{cuts1})--(\ref{eq:cutsqcd}) and the requirement that $p_T^W < 250$ GeV
or $p_T^Z < 100$ GeV -- that is, we modified only the cut (\ref{cutano}) which
is intended to enhance the signal. This choice of the calibration region has
the virtue of preserving the requirements on the jets, and consequently, not
affecting significantly the veto survival probability.

\begin{figure}[!t]
\centerline{\psfig{figure=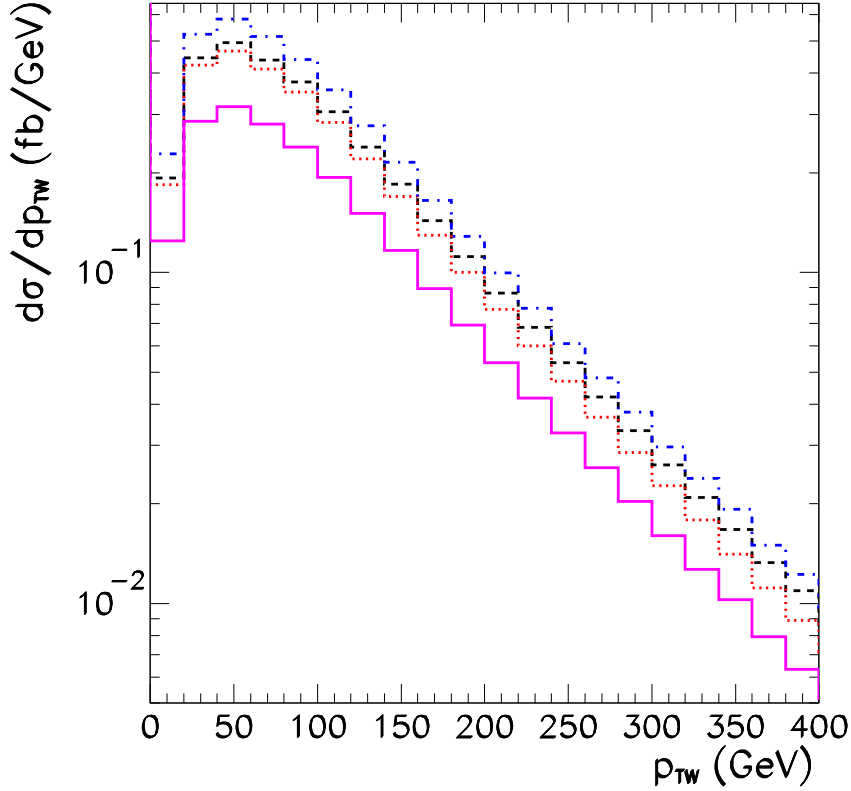,width=5in}} 
\caption{$W^+$ transverse momentum
  distribution for different choices of the renormalization and
  factorization scales after cuts (\ref{cuts1})--(\ref{eq:cutsqcd}).
  The dashed line stands for $\mu_F^0 = \mu_R^0 = \sqrt{(p_{Tj_1}^2 +
    p_{Tj_2}^2)/2}$ and the dash-dotted line represents $\mu_R^0 =
  \sqrt{(p_{Tj_1}^2+p_{Tj_2}^2)/2}$ and $\mu_F^0 = \sqrt{\hat s}$
  where $\hat s$ is the squared parton center--of--mass energy. Our
  default choice $\alpha_s^2(\mu_R^0) = \alpha_s(p_{Tj_1})~
  \alpha_s(p_{Tj_2})$ and $\mu_F^0 = \hbox{min}(p^2_{Tj1}, ~
  p^2_{Tj2},~ p^2_{TW})$ is represented by the solid line while the
  dotted curve stands for $\mu_R^0 = \sqrt{p_{Tj1}~ p_{Tj2}}$ and
  $\mu^0_F = (p_{Tj1}~ p_{Tj2}~ E^2_W)^{1/4}$ where $E^2_W = p^2_{TW}
  + m^2_W$.  }
\label{fig:extra}
\end{figure}

Shown in Fig.\ \ref{fig:extra} is $d\sigma/dp_T^{W^+}$ for four different
choices of the renormalization and factorization scales:
\begin{itemize}
  
\item [{\bf C1}] $\mu_F^0 = \mu_R^0=\sqrt{(p_{Tj_1}^2+p_{Tj_2}^2)/2}$
  (dashed line);
  
\item [{\bf C2}] $\mu_R^0 = \sqrt{(p_{Tj_1}^2+p_{Tj_2}^2)/2}$ and
  $\mu_F^0 = \sqrt{\hat s}$ (dash-dotted line) where $\hat s$ is the
  squared parton center--of--mass energy;
  
\item [{\bf C3}] $\mu_R^0 = \sqrt{p_{Tj1}~ p_{Tj2}}$ and $\mu^0_F =
  (p_{Tj1}~ p_{Tj2}~ E^2_W)^{1/4}$ (dotted curve) where $E^2_W =
  p^2_{TW} + m^2_W$;
  
\item [{\bf C4}] our default choice $\alpha_s^2(\mu_R^0) =
  \alpha_s(p_{Tj_1})~\alpha_s(p_{Tj_2})$ and $\mu_F^0 =
  \hbox{min}(p^2_{Tj1},~ p^2_{Tj2},~ p^2_{TW})$ (solid line).

\end{itemize}
In this figure we have added the electroweak and QCD contributions taking into
account the corresponding veto survival probabilities.

At present we only have leading order (LO) calculations of the $Wjj$ and $Zjj$
QCD backgrounds available. Due to the small difference in weak boson mass, as
compared to {\em e.g.} the large dijet mass required in our event selection,
QCD corrections for these processes are similar and we only present the
$p_T^{W^+}$ distribution for the $W^+jj$ background in the following but in
our analysis we have evaluated the uncertainties for the processes $W^+jj$,
$W^-jj$, and $Zjj$. As we can see from figure \ref{fig:extra}, the
normalization of the background changes by up to a factor of 2 between these
choices.  Moreover, another variation by a factor of 1.5 is obtained by
changing individual renormalization scales between $\mu_R = \mu_R^0/10$ and
$\mu_R = 10\mu_R^0$.  However, while the normalization of the $W^+jj$ cross
section changes drastically, the shape of the $p_T^{W^+}$ distribution is
essentially unaffected.

\begin{figure}[!t]
\centerline{\psfig{figure=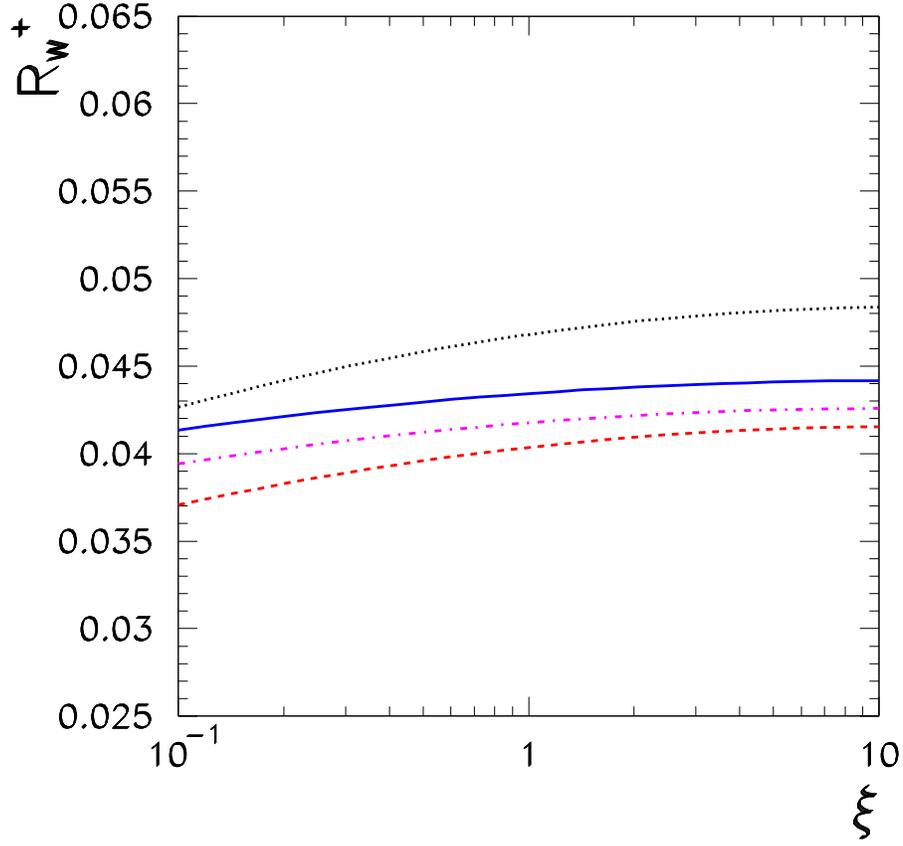,width=5in}} 
\caption{The ratio $R_{W^+}$  is shown as a function of
  $\xi$, where $\mu_R=\xi\mu_R^0$. The lines follow the same
  conventions as in Fig.\ \ref{fig:extra}.  }
\label{fig:scale}
\end{figure}

As a measure of shape changes we study the ratio of the cross sections in the
signal region and the calibration region ($R_V$ for $V= W^\pm,~Z$) as a
function of $\xi$, the scale factor for the four different renormalization
scale choices $\mu_R = \xi\mu_R^0$ listed above. For example for $W^+$ we
define the ratio of the cross sections obtained imposing that $p_T^{W^+}> 300$
GeV and $p_T^{W^+} < 250$ GeV,
\begin{equation}
  R_{W^+} = \frac{\sigma(p_T^{W^+} > 300 \hbox{ GeV})}
   {\sigma(p_T^{W^+} < 250 \hbox{ GeV})}\;\; ,
\label{eq:R1}
\end{equation}
where in the evaluation of these ratios we have added the electroweak and QCD
contributions taking into account the corresponding veto survival
probabilities.  The $\xi$ dependence shown in Fig.~\ref{fig:scale} is small
for individual choices of $\mu_R^0$, being smaller than the differences
between the four basic scales $\mu_R^0$. From this figure it is clear that the
extrapolation uncertainty in this case is rather small ($\simeq 7$\%) despite
the use of a LO QCD calculation.  Certainly, a NLO calculation, which
hopefully will be available by the time the experiment is performed, should
reduce this shape uncertainty.

Altogether the total expected uncertainty in the estimated number of
background events has two sources: the {\sl theoretical} uncertainty
associated to the extrapolations from the calibration region
($\delta_{\text{bck,th}}$) and the {\sl statistical} error associated to the
determination of the background cross section in the calibration region
($\delta_{\text{bck,stat}}$).  Table \ref{tab:err} exhibits our results for
these errors as relative uncertainties assuming an integrated luminosity of
100 fb$^{-1}$.  As we can see from this table, the {\sl theoretical}
extrapolation error is the largest uncertainty.

\section{Results and discussion}

In order to extract the attainable limits on the anomalous TGV's we
assumed an integrated luminosity of $100$ fb$^{-1}$ and that the
observed number of events $Vjj$ with $V= W^\pm,~Z$ is compatible with
the SM expectations for the choice {\bf C4} of the renormalization and
factorization scales both in the signal
($N^{\text{S}}_{\text{V,data}}$) and in the calibration ($
N^\text{C}_{\text{V,data}}$) regions, {\em i.e.}
\begin{eqnarray}
N^{\text{S}}_{\text{V, data}}= N^{\text{S}}_{\text{V,SM,C4}} 
&\;\;\;\;\;\hbox{and}\;\;\;\;\; & 
N^{\text{C}}_{\text{V,data}}=N^{\text{C}}_{\text{V,SM,C4}} \;\; .
\end{eqnarray}
We also assumed no charge discrimination, and consequently, we combined
$W^+jj$ and $W^-jj$ events.


The anomalous TGV's manifest themselves as a difference between the number of
observed events and the number of background events estimated from the
extrapolation of the background measured in the calibration region ($
N_{\text{V,back}}^{\text{S}}$), that is,
\[
     N^{\text{S}}_{\text{V, data}} - N_{\text{V,back}}^{\text{S}} \;\; ,
\] 
where $N_{\text{V, back}}^{\text{S}}= R_V \, N^{\text{C}}_{\text{V,data}}$.
The statistical error of the number of anomalous events is
\begin{equation}
   \sigma_{\text{stat}}^2 = N_{\text{V,data}}^{\text{S}}
     + (R_V\, N_{\text{V,data}}^\text{C} \delta_{\text{bck,stat}})^2
\end{equation}
where the first term is the statistical error of the measured number of events
in the signal region and the second term is the error in the determination of
the background in the signal region due to the statistical error of the
background measurement in the calibration region, $\delta_{\text{bck,stat}}$.
Both errors can be assumed to be gaussian and we combine then in quadrature.

%
%

So at 95\% CL we can impose a limit on the anomalous couplings from the
condition
\begin{equation}
   |N_{V}(g_{\text{ano}})-N_{\text{V,data}}^{\text{S}} +
    N_{\text{V, back}}^{\text{S}}|
   =  |N_{V}(g_{\text{ano}})-N_{\text{V,data}}^{\text{S}} 
      +R_V  N^{\text{C}}_{\text{V,data}}|
     \leq 1.96\, \sigma_{\text{stat}}  \;\; ,
\label{eq:bound}
\end{equation}
where $N_{V}(g_{\text{ano}})$ stands for the expected number of anomalous
events that can be inferred using Eq.~(\ref{def:sigma}).  From
Eq.~(\ref{eq:bound}) it is clear that the extracted bound on the anomalous
couplings depends on what we assume for the range of $R_V$ compatible with the
measured background in the calibration region. In other words, the constraints
depend on how much of the estimated range for the number of background events
due to the extrapolation uncertainty will still be allowed once the
measurement in the calibration region is available.

As a benchmark we first evaluate the attainable 95\% CL constraints on the
TGV's neglecting the extrapolation uncertainty, that is, for
$R_V=N_{\text{V,data}}^{\text{S}}/N^{\text{C}}_{\text{V,data}}$.  In this case
and assuming that only one TGV is non--vanishing we get
\begin{eqnarray}
& -0.18\leq \Delta\kappa_\gamma\leq  0.045 
\;\;\;\;\;\;\;\; , \;\;\;\;\;\;\;\;
& -0.033\leq \lambda_\gamma \leq  0.037  \;\; ,
\nonumber \\
& -0.075\leq \Delta g^Z_1\leq   0.023  
\;\;\;\;\;\; , \;\;\;\;\;\;\;\;
& -0.077 \leq \Delta\kappa_Z\leq  0.029  \;\; ,
\label{eq:limits0}\\
& -0.021\leq \lambda_Z\leq  0.027 
\;\;\;\;\;\;\;\; , \;\;\;\;\;\;\;\;
& -0.12\leq g^Z_5\leq  0.10 \;\; .
\nonumber 
\end{eqnarray}

We also varied the cuts (\ref{cuts1})--(\ref{cutano}) in order to verify
whether these limits could be improved. Nevertheless, it turns out that these
are the best bounds except for $\kappa_\gamma$, which is better constrained
when we relax the following cuts\footnote{In this case we define the
  calibration region with $M_{jj} > 1200 \hbox{ GeV}$ and $p^W_T < 100 \hbox{
    GeV}$ and determine the corresponding uncertainties.}
\begin{equation}
M_{jj} > 1200 \hbox{ GeV} \;\;\;\;\;\; \hbox{and} \;\;\;\;\;\;
p^W_T > 100 \hbox{ GeV} \;\; ,
\label{eq:relaxcuts}
\end{equation}
leading to
\begin{equation}
 -0.036\leq \Delta\kappa_\gamma\leq  0.031  \;\; .
\label{relax0}
\end{equation}

We learn from these results that the analysis of the total number of events
for single production of electroweak gauge bosons has the potential of
improving the constraints with respect to the pair production for the
couplings $\Delta \kappa_\gamma$ and $\Delta \kappa_Z$; see Table
\ref{tab:outros}.

Next we conservatively estimate the 95\% CL sensitivity limits at the LHC
assuming the largest (or smallest) possible background within the full range
of our presently estimated LO extrapolation uncertainty $R_V=N_{\text{
    V,data}}^{\rm S}/N^{\text{C}}_{\text{V,data}} (1 \pm
\delta_{\text{back,th}})$.  In this case the most conservative 95\% CL bound
can be obtained from
\begin{equation}
|N_{V}(g_{\text{ano}})|\leq 
\left |N_{\text{V,data}}^{\text{S}}\, -\,R_V \, N^{\text{C}}_{\text{V,data}}
\right|_{\text{max}}+ 1.96\,\sigma_{\text{stat}}
\leq N_{\text{V,data}}^{\text{S}}\, \delta_{\text{bck, th}}
+ 1.96\,\sigma_{\text{stat}} \;\; ,
\label{limit2}
\end{equation}
which leads to 
\begin{eqnarray}
& -0.20\leq \Delta\kappa_\gamma\leq  0.065 
\;\;\;\;\;\;\;\; , \;\;\;\;\;\;\;\;
& -0.043\leq \lambda_\gamma \leq  0.046  \;\; ,
\nonumber \\
& -0.097\leq \Delta g^Z_1\leq   0.035  
\;\;\;\;\;\; , \;\;\;\;\;\;\;\;
& -0.089 \leq \Delta\kappa_Z\leq  0.039  \;\; ,
\label{eq:limits1}\\
& -0.027\leq \lambda_Z\leq  0.033 
\;\;\;\;\;\;\;\; , \;\;\;\;\;\;\;\;
& -0.14\leq g^Z_5\leq  0.13 \;\; .
\nonumber 
\end{eqnarray}

Last we estimate the attainable 95\% CL constraints at the LHC assuming that
the extrapolation uncertainty will be reduced by a factor two once NLO
predictions are available and/or the data from the calibration region reduces
the allowed range of background predictions:
\begin{eqnarray}
& -0.066\leq \Delta\kappa_\gamma\leq  0.052  
\;\;\;\;\;\;\;\; , \;\;\;\;\;\;\;\;
& -0.038\leq \lambda_\gamma \leq  0.042  \;\; ,
\nonumber \\
& -0.086\leq \Delta g^Z_1\leq   0.029  
\;\;\;\;\;\; , \;\;\;\;\;\;\;\;
& -0.083 \leq \Delta\kappa_Z\leq  0.034  \;\; ,
\label{eq:limits2}\\
& -0.024\leq \lambda_Z\leq  0.030 
\;\;\;\;\;\;\;\; , \;\;\;\;\;\;\;\;
& -0.13\leq g^Z_5\leq  0.12 \;\; ,
\nonumber 
\end{eqnarray}
where the bounds on $\Delta \kappa_\gamma$ have been obtained with the relaxed
cuts in Eq.~(\ref{eq:relaxcuts}).

In deriving these bounds we have statistically combined the results from
$W^\pm jj$ and $Zjj$ production although the limits originate basically from
$W^\pm jj$ production, having the $Zjj$ process a marginal impact on the
constraints due the the small SM--anomalous interference cross section; see
Table \ref{tab:sigmaz}. The only anomalous couplings for which the $Zjj$
production plays any role is $g^Z_1$.  Moreover, the presence of
non--vanishing interference between the SM and anomalous contributions lead to
bounds that are asymmetrical around zero.

\begin{figure}[!t]
\centerline{\psfig{figure=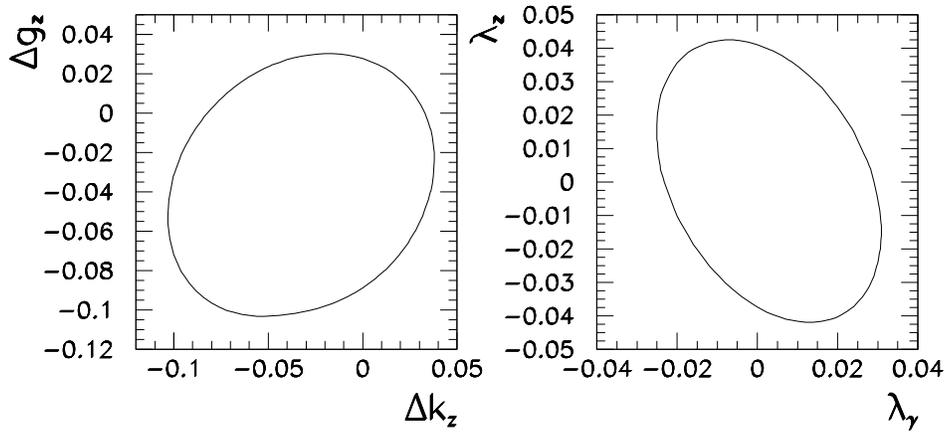,width=5in}} 
\caption{95\% CL allowed region in the planes $(\Delta g_1^Z,\Delta
  \kappa_Z)$ and $(\lambda_Z,\lambda_\gamma)$.  }
\label{fig:contour}
\end{figure}

We can see from Tables \ref{tab:sigmaw+}--\ref{tab:sigmaz} that there are
non-vanishing interference between the different anomalous TGV's
contributions, and consequently, there will be nontrivial correlations between
the bounds on these couplings when more than one coupling is non vanishing. We
depict in Fig.\ \ref{fig:contour} the 95\% CL (2dof) regions in the planes
$(\Delta g_1^Z , \Delta \kappa_Z)$ and $(\lambda_Z , \lambda_\gamma)$. As we
can see from this figure there is correlation between $(\Delta g_1^Z , \Delta
\kappa_Z)$ and anti-correlation between $(\lambda_Z , \lambda_\gamma)$.

In the framework of effective lagrangians exhibiting a linear realization of
the $SU(2)_X \otimes U(1)_Y$ symmetry, the anomalous TGV's satisfy the
relations (\ref{lin:1})--(\ref{lin:3}). Assuming these constraints among the
anomalous interactions and $f_B = f_W$, the potential LHC 95\% CL bounds are
\begin{eqnarray}
&  -0.052~ (-0.034) \leq \Delta\kappa_\gamma\leq  0.040~(0.028)  \;\; ,
\nonumber \\
&  -0.019~(-0.017) \leq \lambda_{\gamma(Z)}  \leq  0.023~(0.021) \;\; ,
\nonumber \\
&  -0.097~(-0.090) \leq \Delta g_1^Z\leq   0.019~(0.016) \;\; ,
\\
&  -0.052~(-0.049) \leq \Delta\kappa_Z\leq  0.010~(0.0085) \;\; ,
\nonumber
 \label{eq:limHISZ}
\end{eqnarray}
when we use the full (half) LO extrapolation uncertainty and the relaxed cuts
for $\Delta \kappa_\gamma$.  The constraints on $g_5^Z$ are the ones given in
Eqs.\ (\ref{eq:limits1}) and (\ref{eq:limits2}) since $g_5^Z$ is not related
to the other couplings in the HISZ scenario.

The improvement of the constraints on $\lambda_{\gamma(Z)}$ in the HISZ
scenario is easy to understand remembering that the TGV's $\lambda_Z$ and
$\lambda_{\gamma}$ are anti-correlated as shown in Fig.\ \ref{fig:contour}. On
the other hand, the tighter bounds obtained for $\Delta \kappa_\gamma$,
$g_1^Z$, and $\Delta \kappa_Z$ originates from the hypothesis $f_W = f_B$
which relates these couplings, leading to just one independent combination of
them.

We can learn from the above results that the study of the production rates of
single electroweak gauge bosons via WBF at the LHC can lead to
upper bounds on $\Delta\kappa_Z$ that are of the same order of the bounds
derived from the kinematical analysis of pair production of electroweak 
gauge bosons.  In the case of
the anomalous coupling $\Delta \kappa_\gamma$ the upper bound can be slightly
weaker or of the same order as that coming from gauge boson pair production.
On the other hand, the bounds on $\lambda_{Z,\gamma}$ and $g_1^Z$ obtained via
the latter process are more stringent.

\begin{figure}[!t]
\centerline{\psfig{figure=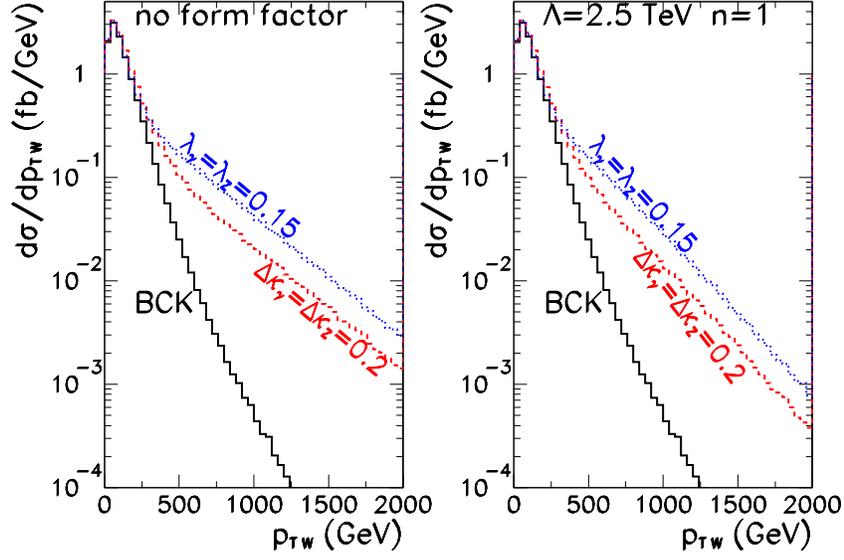,width=5in}} 
\caption{Transverse momentum distribution for the $W^+$ produced
  in $W^+jj$ events with $W^+\rightarrow l^+\nu_l$ at LHC after applying the
  cuts (\ref{cuts1})--(\ref{cutano}). The left panel shows the distribution
  for the background and anomalous TGV signal without the effect of the form
  factor (\ref{formfactor}).  On the right panel this form factor has been
  included for $n=1$ and $\Lambda = 2.5$ TeV.}
\label{fig:distptw}
\end{figure}

\begin{figure}[!t]
\centerline{\psfig{figure=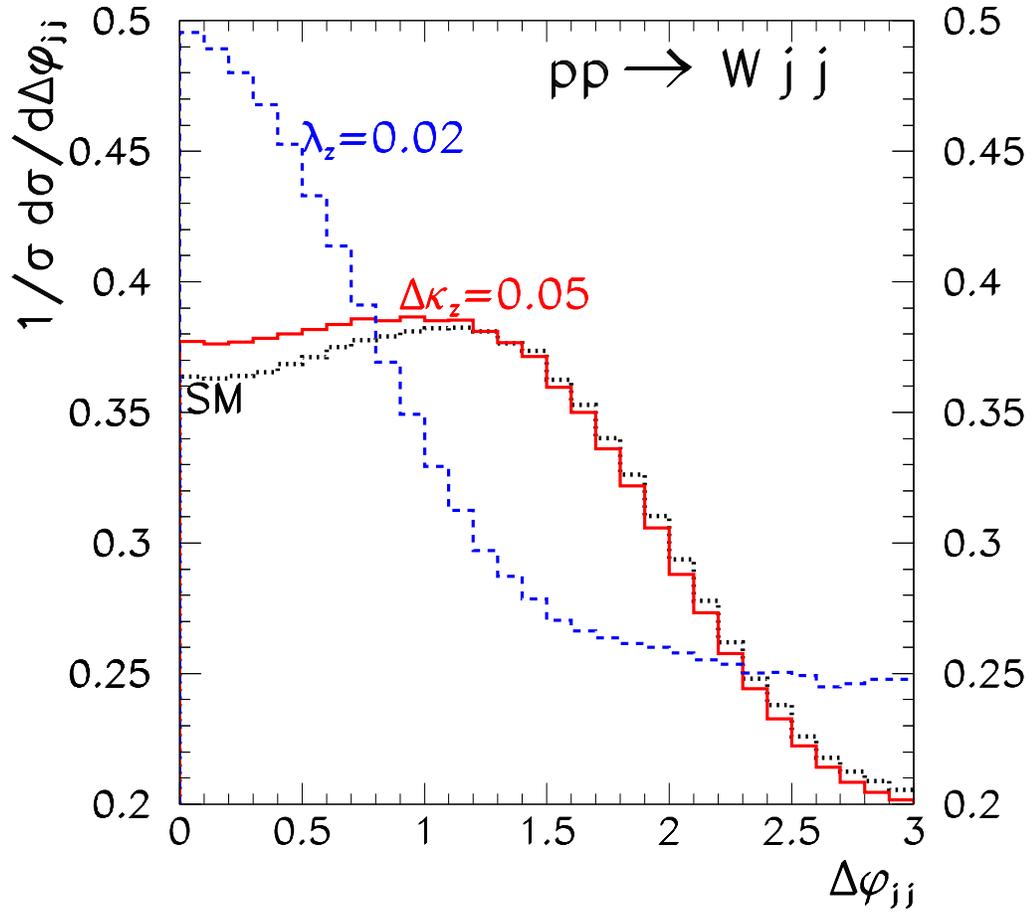,width=5in}} 
\caption{Normalized $\Delta \varphi_{jj}$ distribution for $W^+ jj$. The
  dotted line stands for the SM result while the dashed (full) line
  stands for the resulting assuming $\lambda_Z = 0.02$ ($\Delta
  \kappa_Z =0.05$).  }
\label{fig:phijj}
\end{figure}

There is still room for further improvements in our analyses, which are beyond
the scope of this work. First, the importance of the single $Z$ production can
be enlarged by considering the invisible decay of the $Z$ into neutrino pairs.
This process can certainly be extracted from the backgrounds analogously to
the case of invisibly decaying Higgs bosons~\cite{oedz}. Second, the use of
higher order QCD calculations might lead to lower extrapolation errors from
the calibration regions as illustrated above. Probably the most significant
improvement will be the use of kinematical distributions to constrain the
anomalous TGV's in analogy with what is done in the analysis of 
electroweak gauge boson pair production~\cite{tri@lhc}.  Notice that with 
the cuts proposed the final data sample
contains $\sim$ 770 electroweak $W^\pm$ events and $\sim$ 480 $Z$ events for
an integrated luminosity of ${\cal L}=100$ fb$^{-1}$ allowing for
statistically meaningful binning.  In general the anomalous couplings can be
better constrained by fitting the electroweak vector boson transverse momentum
distribution; see Fig.\ \ref{fig:distptw}.  The anomalous couplings
$\lambda_{Z,\gamma}$ can be also better determined by studying the angle in
the transverse plane between the tagging jets ($\Delta \varphi_{jj}$). In
fact, Fig.\ \ref{fig:phijj} shows that the shape of the $\Delta \varphi_{jj}$
distribution is quite different for these couplings when compared to the SM
predictions or the other anomalous contributions. Therefore should an event
excess be found, this distribution can be used to discriminate among the
different couplings.

In summary we have shown that the single production of electroweak gauge
bosons via the weak boson fusion processes at LHC can provide stringent tests
on the deviations of the TGV's from the standard model prediction. This is
possible because the QCD background can be efficiently reduced by exploiting
the difference in several kinematical distributions. Furthermore we have shown
that the background can be estimated with good enough precision by
extrapolation from the measured rate in a signal-suppressed calibrating region
of the phase space. Altogether the sensitivity bounds obtained for some of the
anomalous couplings are comparable to those attainable from the study of
electroweak gauge boson pair production. With these results we stress the
importance of studying both type of reactions to obtain maximum information on
the gauge structure of the electroweak interactions.


\acknowledgments 

We thank J. Hobbs, M. Maltoni and R. Zukanovich--Funchal for very useful
discussions and comments.  M.C. Gonzalez-Garcia would like to thank IF-USP,
while O.~\'Eboli would like to thank Y.I.T.P.  for their hospitality during the
development of most of this work.  This work was supported by Conselho
Nacional de Desenvolvimento Cient\'{\i}fico e Tecnol\'ogico (CNPq), by
Funda\c{c}\~ao de Amparo \`a Pesquisa do Estado de S\~ao Paulo (FAPESP), by
Programa de Apoio a N\'ucleos de Excel\^encia (PRONEX). MCG-G acknowledges
support from National Science Foundation grant PHY0098527 and Spanish Grants
No FPA-2001-3031 and CTIDIB/2002/24.



\begin{table}
\begin{tabular} {|c||c|c|c|}
\hline
anomalous coupling   &  direct LEP limits  & indirect limits  & pair
production limits at the LHC
\\ \hline
$\Delta \kappa_\gamma$ &  $[-0.105~,~0.069]$ & $[-0.044~,~0.059]$  &
                                                         $[-0.034~,~0.034]$
\\ \hline
$\lambda_\gamma$   & $[-0.059~,~0.026]$  & $[-0.061~,~0.10]$ &
                                                         $[-0.0014~,~0.0014]$
\\ \hline
$g_1^Z$            & $[-0.051~,~0.034]$ & $[-0.051~,~0.0092]$ & 
                                                         $[-0.0038~,~0.0038]$
\\ \hline
$\Delta \kappa_Z$ & $[-0.040~,~0.046]$ & $[-0.050~,~0.0039]$ & 
                                                         $[-0.040~,~0.040]$
\\ \hline
$\lambda_Z$ & $[-0.059~,~0.026]$ &$[-0.061~,~0.10]$ & $[-0.0028~,~0.0028]$
\\ \hline
$g_5^Z$ & $[-0.95~,~0.079]$ & $[-0.085~,~0.049]$ & ---
\\\hline
\end{tabular}
\caption{95\% CL limits on the anomalous couplings emanating from
 direct measurements at LEP2 and from loop contributions to the precision 
measurements at LEP I
~\cite{LEPI},
 assuming the HISZ scenario  $\Delta \kappa_\gamma = \frac{c_W^2}{s_W^2}
\left( \Delta g_1^Z - \Delta \kappa_Z  \right)$ and $\lambda_Z =
\lambda_\gamma$. We also present the expected 95\% CL bounds at
 the LHC obtained through the pair production of electroweak gauge bosons.
The entry marked as --- has not been evaluated in the 
literature.}
\label{tab:outros}
\end{table}


\begin{table}
\begin{tabular} {|c||cc|c|c|c|c|c|c|}
\hline
& \multicolumn{2}{c|}{SM} & \multicolumn{6}{c|}{anomalous} \\
\hline
$\sigma_{W^+}\times P_{\rm surv} \times {\rm eff}$ 
(fb) 
& QCD & EW &
$\Delta\kappa_\gamma$ & $\lambda_\gamma$ &
$\Delta g_1^Z$  & $\Delta\kappa_Z$ & $\lambda_Z$ & $g_5^Z$ \\
\hline\hline
SM &   0.71 & 4.70 &
   6.5 &    -1.2 &    12.0 &     8.8 &    -4.4 &     1.0 \\\hline
$\Delta\kappa_\gamma$ &    
\multicolumn{2}{c|}{---} &
 49.0 &    -7.5 &   -17.1 &    70.9 &    -4.3 &    -1.6 \\ \hline
$\lambda_\gamma$  &   \multicolumn{2}{c|}{---} 
&     --- &
323.3 &     8.8 &    -4.3 &   354.0 &    -1.3 \\ \hline
$\Delta g_1^Z$   &   \multicolumn{2}{c|}{---} 
&     --- &     --- &
161.1 &   -70.7 &    35.5 &    11.9 \\ \hline
 $\Delta\kappa_Z$ &    \multicolumn{2}{c|}{---} 
&     --- &     --- &     --- &
 180.7 &   -16.0 &    -2.0 \\ \hline
$\lambda_Z$   &  \multicolumn{2}{c|}{---} 
&     --- &     --- &     --- &     --- &  735.1 &    -6.5 \\ \hline
$g_5^Z$  &   \multicolumn{2}{c|}{---} 
&     --- &     --- &     --- &     --- &     --- &    31.5 
\\ \hline
\end{tabular}
\caption{
 SM, anomalous and interference terms as defined in Eq.~(\ref{def:sigma}) 
for $W^+$ production after applying the cuts  (\ref{cuts1})--(\ref{cutano}). 
}
\label{tab:sigmaw+}
\end{table}


\begin{table}
\begin{tabular} {|c||cc|c|c|c|c|c|c|}
\hline
& \multicolumn{2}{c|}{SM} & \multicolumn{6}{c|}{anomalous} \\
\hline
$\sigma_{W^-}\times P_{\rm surv} \times {\rm eff}$ 
(fb) 
& QCD & EW &
$\Delta\kappa_\gamma$ & $\lambda_\gamma$ &
$\Delta g_1^Z$  & $\Delta\kappa_Z$ & $\lambda_Z$ & $g_5^Z$ \\
\hline\hline
SM & 
  0.30&   1.98 &  
2.9 &    -0.4 &     4.8 &     3.8 &    -1.7 &    -0.5 \\\hline
 $\Delta\kappa_\gamma$ &    
\multicolumn{2}{c|}{---} &    
20.5 &    -3.2 &    -7.2 &    29.1 &    -1.4 &     1.1 \\\hline
 $\lambda_\gamma$  &   \multicolumn{2}{c|}{---} 
 &    --- &   
 127.8 &     3.5 &    -1.4 &   132.2 &    -0.8 \\\hline
$\Delta g_1^Z$   &   \multicolumn{2}{c|}{---} 
&    ---&    ---&    
 64.0 &   -29.6 &    15.2 &    -2.6 \\\hline
 $\Delta\kappa_Z$ &    \multicolumn{2}{c|}{---} 
&    ---&    ---&    ---&    
74.4 &    -6.3 &     3.6 \\\hline
$\lambda_Z$   &  \multicolumn{2}{c|}{---} 
&    ---&    ---&    ---&    ---&    281.2 &    -1.7 \\\hline
$g_5^Z$  &   \multicolumn{2}{c|}{---} 
&    ---&    ---&    ---&    ---&    ---&   13.2 
\\\hline
\end{tabular}
\caption{
 SM, anomalous and interference terms as defined in Eq.~(\ref{def:sigma}) 
for $W^-$ production after applying the cuts
  (\ref{cuts1})--(\ref{cutano}). 
 }
\label{tab:sigmaw-}
\end{table}


\begin{table}
\begin{tabular} {|c||cc|c|c|c|c|}
\hline
& \multicolumn{2}{c|}{SM} & \multicolumn{4}{c|}{anomalous} \\
\hline
$\sigma_{Z}\times P_{\rm surv} \times {\rm eff}$ 
(fb) 
& QCD & EW &
$\Delta g_1^Z$  & $\Delta\kappa_Z$ & $\lambda_Z$ & $g_5^Z$ \\
\hline\hline
SM &  1.37 &   3.40&
 9.0 &     0.0 &     0.6 &     0.0 \\\hline
$\Delta g_1^Z$   &   \multicolumn{2}{c|}{---} &
     22.7 &     0.2 &    -8.7 &     0.0 \\\hline
 $\Delta\kappa_Z$ &    \multicolumn{2}{c|}{---}&  
--- &  
8.3 &    11.1 &     0.0 \\\hline
$\lambda_Z$   &  \multicolumn{2}{c|}{---} & 
    --- &     --- &    
  52.8 &     0.0 \\\hline
$g_5^Z$  &   \multicolumn{2}{c|}{---} &
    --- &     --- &     --- &     13.9 \\
\hline
\end{tabular}
\caption{
 SM, anomalous and interference terms as defined in Eq.~(\ref{def:sigma}) 
for $Z$ production after applying the cuts  (\ref{cuts1})--(\ref{cutano}). 
 }
\label{tab:sigmaz}
\end{table}


\begin{table}
\begin{tabular} {|c||c|c|}
\hline
Final state  &   $\delta_{bck,stat}$ (\%)   &  $\delta_{bck,th}$ (\%)\\
\hline \hline
$W^+jj$      &   0.90 &  7.4  
\\
\hline
$W^-jj$      &   1.3 &  5.7 
\\
\hline
$W^\pm jj$      &   0.74 & 4.4 
\\
\hline
$Zjj$        &   5.4  & 2.6  
\\
\hline
\end{tabular}
\caption{Uncertainties of the background estimate. The statistical error
was estimated assuming an integrated luminosity of $100 \hbox{
 fb}^{-1}$. }
\label{tab:err}
\end{table}


\end{document}